\documentclass[aps,twocolumn,nopacs,floats,superscriptaddress]{revtex4-2}
\usepackage{graphicx}
\usepackage{dcolumn}
\usepackage{bm}
\usepackage{amsmath}
\usepackage{amssymb}
\usepackage{color}

\usepackage{xcolor}

\definecolor{cardinal}{rgb}{0.6,0,0}
\definecolor{darkgreen}{rgb}{0,0.4,0}
\definecolor{golden}{rgb}{0.92, 0.7, 0}
\definecolor{midnight}{rgb}{0, 0, 0.5}
\definecolor{darkblue}{rgb}{0, 0, 0.7}

\def\he4{$^4$He}
\def\hel3{$^3$He}
\def\Am3{\AA$^{-3}$}
\def\beq{\begin{equation}}
\def\eeq{\end{equation}}

\newcommand{\cH}{{\mathcal H}}

\newcommand{\be}{\begin{equation}}
\newcommand{\ee}{\end{equation}}
\newcommand{\bea}{\begin{eqnarray}}
\newcommand{\eea}{\end{eqnarray}}
\newcommand{\bse}{\begin{subequations}}
\newcommand{\ese}{\end{subequations}}
\newcommand{\dr}{\mathrm{d}}

\begin{document}

\author{Viktor Berger}
\affiliation{Department of Physics, University of Massachusetts, Amherst, MA 01003, USA}

\author{Nikolay Prokof'ev}
\affiliation{Department of Physics, University of Massachusetts, Amherst, MA 01003, USA}

\author{Boris Svistunov}
\affiliation{Department of Physics, University of Massachusetts, Amherst, MA 01003, USA}

\title{Universal
Low-temperature Depletion of Superfluid Density in the Absence of Galilean Symmetry
}

\begin{abstract}
Landau theory of superfluidity associates low-temperature flow of the normal component with the phonon wind. This picture does not apply to superfluids in which Galilean invariance is broken either by disorder, porous media, or lattice potential,
and the phonon wind is no longer solely responsible for depletion of the superfluid component. 
Based on Popov's hydrodynamic action with anharmonic terms, 
we present a general theory for low-temperature ($T$) dependence of the superfluid stiffness, which reproduces Landau result as a special case when several parameters of the hydrodynamic action are fixed by Galilean invariance, and validate it with numerical simulations of interacting lattice bosons. 
In a broader context, our approach reveals universal low-temperature thermodynamics of superfluids with an intrinsic connection between finite-$T$ and finite-size ($L$) effects implying universal scaling, $T^{d+1}$ and $1/L^{d+1}$, respectively, for a large class of thermodynamic quantities. We discuss the experimental detection of this law, and compare our prediction to the existing literature.
\end{abstract}

\maketitle


\textit{Introduction.} 
The concept of superfluid phase \cite{Anderson_1966} is central for the modern theory of superfluidity. It is in terms of the superfluid phase field that one accounts for all universal properties of superfluids such as quantization of the superfluid velocity circulation \cite{Onsager_1949,Feynman_1955}, isolated vortexes and vortex arrays \cite{Feynman_1955, Andronikashvili1966}, 
AC Josephson effect \cite{Josephson}, and universality classes of superfluid transitions; for a review, see, e.g., Ref.~\cite{book}. In a sharp contrast, the foundational Landau theory of superfluidity \cite{Landau_1941} directly associates the universal low-temperature normal flow with the phonon wind---a ballistic flow of thermally equilibrated phonon subsystem past superfluid vacuum, thereby creating an impression that full understanding of the universal mechanism of low-temperature depletion of the superfluid density can be achieved without invoking the notion of superfluid phase.

The key questions then are how the problem is solved using the concept of superfluid phase and whether Landau-type arguments can be generalized to systems with continuous translation symmetry being broken either by periodic or disordered external potential (we answer this question \textit{negatively}). 
Important examples of such systems include ultracold atoms in optical lattices, supersolids, \he4 in porous media \cite{Mehl1968}, \he4 films on substrates \cite{Boninsegni2010}, as well as superfluid and supersolid excitons in double-layer structures \cite{conti2023chester, rios2018evidence}. The two-fluid picture
needs to be seriously revised in all such cases because at $T=0$ only the superfluid component is flowing while the non-zero normal component is pinned by external potential. Nevertheless, a well-defined long-wave phonon
subsystem exists even in this case. The hydrodynamic Hamiltonian describing the phonon subsystem is translation invariant, and one may
think that the temperature dependence of depletion can be explained by
the phonon wind. 

In Landau theory, normal flow and phonon wind are synonymous, yielding a straightforward way of calculating the depletion of $\rho_s$ at low temperature: By Galilean invariance, the
phonon wind in the reference frame moving with the superfluid velocity exactly corresponds to the pure superflow in the reference frame in which the phonon subsystem is at rest.  Therefore, the momentum density of the equilibrium phonon wind as a function of its (infinitesimal) velocity immediately yields the normal density $\rho_n$ and
$\Delta \rho_s =-\rho_n$:
\be
\Delta \rho_s=-\frac{I_d}{d}\frac{d+1}{mc(c\beta)^{(d+1)}},
\quad I_d=\int\frac{\mathrm{d}^dq}{(2\pi)^d}\frac{q}{e^q-1}, 
\label{Landau}
\ee
where $m$ is the particle mass, $c$ is the sound velocity, and 
$\beta = 1/ T$. 
To ``quantify" the power and limitations of Landau theory, observe that this derivation
rests on the Galilean relation between the mass density flux, ${\bf j}_m$, 
and momentum density, ${\bf p}$:
\be
{\bf j}_m \, =\, {\bf p} \qquad \qquad \mbox{(Galilean system)} \, .
\label{j_to_p}
\ee
This relation is true no matter whether the system is in the low-temperature regime, where $\rho_n$ corresponds to a dilute gas of elementary excitations (phonons), or it is in a strongly correlated regime of critical fluctuations on approach to the critical temperature. 
However, only at sufficiently low temperature is Eq.~(\ref{j_to_p})
practically useful for obtaining $\Delta \rho_s$ through the calculation
of the momentum density of non-interacting elementary excitations subject to 
Gibbs distribution. 

In the absence of continuous translation invariance, the phonon wind will always decay and the genuinely equilibrium superflow regime implies the absence of phonon wind! On the one hand, it would be na\"ive---and, as we will argue, fundamentally wrong---to directly relate the phonon wind momentum as a function of velocity to the depletion of superfluid density.
On the other hand, it is reasonable to expect (given that at $T\to 0$ phonons are the only elementary excitations described by the universal hydrodynamic Hamiltonian) that the temperature dependence of depletion is still universal and even reminiscent of the Landau result, but is controlled by several parameters of the hydrodynamic Hamiltonian
that can take rather arbitrary values in a general case. 

On the qualitative side, the general theory of depletion requires going beyond the harmonic effective Hamiltonian/action. While early treatments of anharmonic terms in the Hamiltonian, 
$\mathcal{H}$, focused on the kinetics of elementary excitations \cite{khalatnikov2018introduction}, our statistical treatment also considers higher-order fluctuations of the superfluid phase field. Reproducing Landau result---originally based on a purely harmonic treatment bootstrapped by the relation (\ref{j_to_p})---using anharmonic terms in $\mathcal{H}$ is both instructive and immediately reveals how special the Galilean invariant systems are. In the general case, even the very notion of
temperature-induced ``depletion" becomes questionable because $\rho_s$ may increase
with temperature, i.e. the theory allows depletion to be ``negative"!

In this Letter, we find that the low-temperature change of superfluid stiffness,
$\Delta n_s(n,\beta) =  n_s(n,\beta)-n_s(n,\infty)$, of a $d$-dimensional superfluid 
at fixed total number density $n$ (canonical ensemble) is given by the formula
($n_s$ and $\rho_s$ are simply related by $\rho_s= m n_s$)
\be\label{eqn:delta_ns_final}
\Delta n_s(n,\beta)=-\frac{I_d}{d}\frac{\nu^2(d+1)-d\gamma n_s -(d+2)\sigma/\varkappa}{c(c\beta)^{(d+1)}}\,,
\ee
\be\label{eqn:nu}
\nu=\frac{\dr n_s}{\dr n}\,, \quad \gamma=\frac{1}{2}\frac{\dr^2n_s}{\dr n^2}\,, \quad \sigma = 2\frac{\partial^2 \mathcal{E}(n,k_0^2)}{\partial(k_0^2)^2}\bigg|_{k_0=0}\, ,
\ee
where all parameters refer to various ground-state properties such as 
$c=\sqrt{n_s/\varkappa}$, 
compressibility $\varkappa = d n / d \mu$, 
chemical potential $\mu$, 
energy density ${\cal E}(n, k_0^2)$ 
(as a function of $n$ and the square of the superflow wavevector $k_0$), 
and its derivatives: 
superfluid stiffness $n_s \equiv n_s(T=0) = 2 \partial  {\cal E}(n, k_0^2)/\partial (k_0^2) $, 
$\nu$, $\gamma$, and $\sigma$.
Unless stated otherwise, we will work in units where Planck's constant is set equal to unity. 
[The superfluid stiffness and superflow wavevector are the most convenient quantities to discuss the general case, especially lattice models.] 

Our result, Eqs.~(\ref{eqn:delta_ns_final})--(\ref{eqn:nu}), 
differs from the original Landau formula (\ref{Landau}) by a factor. 
In a Galilean system, both $\gamma$  and $\sigma$ have to be identically zero and $\nu=1/m$; this is how Eqs.~(\ref{eqn:delta_ns_final})--(\ref{eqn:nu}) recover the Landau expression, while revealing its quantitative and conceptual deficiency in a general case. The relations $\nu=1/m$ and $\gamma=0$ follow from the simple fact that in a Galilean system, the superfluid density is equal to the normal density at zero temperature, while $\sigma=0$ is enforced by the transformation of the system energy under a Galilean boost, see the discussion under Eq.~(\ref{eqn:H_hydro_0}). On the technical side, we derive Eqs.~(\ref{eqn:delta_ns_final})--(\ref{eqn:nu}) starting from Popov's $(d+1)$-dimensional hydrodynamic action \cite{Popov} based on the above-mentioned parameters and without invoking reference frame transformations. 

Strictly speaking, both Landau theory and our generalized description
may go beyond Eqs.~(\ref{eqn:delta_ns_final})--(\ref{eqn:nu}) corresponding to the asymptotic regime when only the acoustic part of the phonon branch is excited. 
On the one hand, as long as the gas of elementary excitations
may be treated as non-interacting, Landau formula remains valid
regardless of the phonon dispersion relation, e.g. it correctly captures
the roton contribution in \he4. [Achieving the same goal in the general case is far more challenging because one needs to include a large number of higher-order aharmonic terms.]  
On the other hand, experimental observation of universal scaling $1/c(\beta c)^{d+1}$ is extremely challenging even in bulk \he4 under ambient pressure. 
Since roton contribution to the system thermodynamics cannot be neglected already at temperature $T>0.5~K$ this power-law scaling is limited to low temperature $T<0.5~K$ 
when thermal depletion of $n_s$ drops below experimental resolution. The same limitation applies to \he4 experiments in porous/powder media where the typical pore size significantly exceeds the roton wavelength.

Below we present the theory of low-temperature depletion of the superfluid stiffness, which yields Eq.~(\ref{eqn:delta_ns_final}) when applied to the case of an infinite system  without Galilean symmetry. Based on our predictions, we discuss 
experiments on \he4 in disordered media. While there exists experimental evidence in support of our theory, we conclude that further studies are needed to determine whether the observed scaling is truly representative of the asymptotic low-temperature regime.

In a companion paper \cite{companion}, we present all the technical details of our theory
and also explain how to produce explicit formulas for the depletion of $n_s$ in the most general case, including finite-size effects and the grand-canonical description. 
The companion paper also discusses the mathematically similar problem of finite-size corrections to the superfluid stiffness in finite-temperature classical-field U(1) systems, as well as provides (extra) supportive numeric results, for both classical and quantum systems.

\textit{The theory.} There are two ways to calculate $n_s$ within Popov's formalism. 
One can either employ the hydrodynamic Hamiltonian 
or, equivalently, use the Lagrangian framework. 
For reasons explained below, we proceed here with the former approach.

By definition, the superfluid stiffness relates the supercurrent density $\langle\mathbf{j}\rangle$ to the infinitesimal superflow wavevector $\mathbf{k_0}$
(constant phase gradient):
\be\label{eqn:ns_def}
\langle\mathbf{j}\rangle=n_s\mathbf{k}_0 \qquad \qquad (k_0\rightarrow0)\, .
\ee
In the long-wave limit, current density is given by \cite{book}
\be\label{eqn:current_def}
\mathbf{j} = \frac{\partial\mathcal{H}(\eta, \nabla\Phi)}{\partial(\nabla\Phi)} \, ,
\ee
where ${\cal H}(\eta,\nabla\Phi)$ is the hydrodynamic Hamiltonian expressed in terms of the phase gradient $\nabla\Phi$ and the number density fluctuations $\eta$ around the expectation value $n$ at equilibrium. It can be derived from a perturbative expansion of the ground-state energy $\mathcal{E}(n,k_0^2)$ in $\eta$ and $k_0^2$, with subsequent replacement  $k_0^2\to(\nabla\Phi)^2$. For the purpose of computing the leading contribution to $\Delta n_s$, the relevant terms in the hydrodynamic Hamiltonian are
\begin{align}\label{eqn:H_hydro_0}
\begin{split}
\mathcal{H}(\eta,\nabla\Phi) & = \frac{n_s^{(0)}}{2}(\nabla\Phi)^2+\frac{\eta^2}{2\varkappa}+\frac{\nu\eta}{2}(\nabla\Phi)^2 \\
& + \frac{\gamma\eta^2}{2}(\nabla\Phi)^2+\frac{\sigma}{4}(\nabla\Phi)^4\,.
\end{split}
\end{align}
At this point it becomes clear why we must have $\sigma=0$ in the Galilean system: The transformation law for the energy forbids terms of order $k_0^4$ and higher, implying that corresponding terms in the hydrodynamic Hamiltonian must vanish.

By default, Popov's hydrodynamic formalism works in the grand canonical ensemble. As such, there will be two contributions to the superfluid stiffness; the ``pure" depletion due to thermal excitations, as well as depletion due to temperature-induced change in the total density. For the purpose of generalizing Landau theory, this is a problem since Landau's theory is formulated in the canonical ensemble. This is where the advantages of the hydrodynamic Hamiltonian are revealed. Apart from yielding an explicit expression for the current density through Eq.~(\ref{eqn:current_def}), when expressed in terms of phase gradients and density deviations the Hamiltonian formalism allows one to identify the contributions to $\Delta n_s$ originating from changes in total density. Indeed, when the phase field is separated into fluctuations around a homogeneous infinitesimal superflow (assumed to be in the $\hat{\mathbf{x}}$-direction),
\be\label{eqn:phi_decomp}
\Phi = k_0x+\varphi\, ,
\ee
the $x$-component of the current density becomes (to the first order in $k_0$)
\bea
j_x  =  n_s^{(0)}k_0     &+& k_0[\gamma \eta^2 +  2 \sigma \varphi_x^2 + \sigma (\nabla \varphi)^2 ] + \nu \eta \varphi_x \nonumber \\
 &+&\,  n_s^{(0)} \varphi_x  \,+\, \nu k_0  \, \eta  \,  +\,  \sigma (\nabla \varphi)^2 \varphi_x \, .
\label{j_hydro_new}
\eea
From here, we observe that in the canonical ensemble, the expectation value of $\eta$ is identically equal to zero, meaning that the corresponding term in the second line of Eq.~(\ref{j_hydro_new}) should be omitted. Combining Eqs.~(\ref{eqn:ns_def}) and~(\ref{j_hydro_new}) reveals that the problem of computing superfluid stiffness depletion now has been reduced to computing a finite number of correlation functions of the fields $\eta$ and $\varphi$. For extracting the leading contribution to the depletion, it is sufficient to compute these correlators with the bi-linearized Hamiltonian
\be\label{eqn:H_hydro}
\cH(\eta,\nabla\varphi)\, =\, \frac{n_s^{(0)}}{2}(\nabla\varphi)^2+\frac{\eta^2}{2\varkappa}+k_0\nu\eta\varphi_x\,.
\ee
This implies that the remaining two terms in the second line of Eq.~(\ref{j_hydro_new}) can be omitted as they are based on odd powers of $\varphi$. Straightforward (although lengthy) calculations of the expectation value of the first line in Eq.~(\ref{j_hydro_new}) yields the result~(\ref{eqn:delta_ns_final}). All technical details of these calculations can be found in Ref.~\cite{companion}.

\begin{figure}
    \centering
    \includegraphics[width=0.9\linewidth]{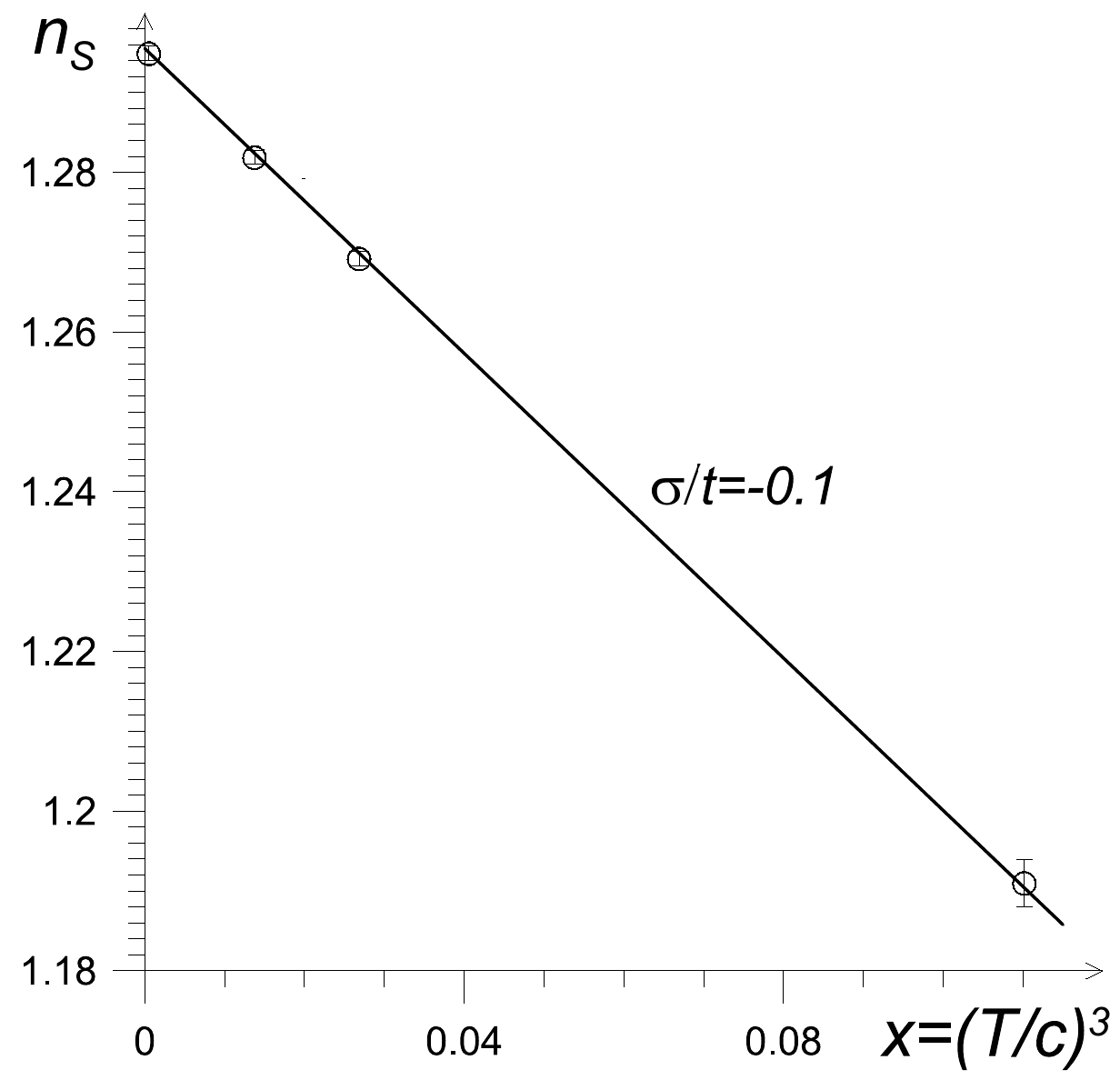}
    \caption{Superfluid stiffness depletion as a function of $x=(cT)^3$ 
    for soft-core bosons in 2D with $U/t=4$ (for system size $L\times L=64\times64$).
    Solid (black) curve is the $f(x) = A + Bx$ theoretical prediction 
    with $A=n_s(T=t/6)$ and $B$ fixed by the measured system parameters according to Eq.~(\ref{eqn:delta_ns_final}). 
    }
    \label{fig:nsT3soft}
\end{figure}

\textit{Numerics.} To verify Eqs.~(\ref{eqn:delta_ns_final})--(\ref{eqn:nu}), we performed Worm Algorithm path-integral Monte Carlo simulations \cite{Worm,PRE} of soft-core bosons on a two-dimensional square lattice at low temperatures and system volume $V=64\times64$. The model is defined by the Bose-Hubbard Hamiltonian
\begin{equation}\label{eqn:bose_hubbard}
    H = -t\sum_{\langle i,j\rangle}(c_i^{\dagger}c_j+\text{H.c.})+U\sum_in_i^2-\mu\sum_in_i \, ,
\end{equation}
where $c_i^{\dagger}$ ($c_i$) is bosonic creation (annihilation) operator
on site $i$.
By varying chemical potential in conjunction with temperature to keep the number density constant, we work in the canonical ensemble, and, thus, 
expect the superfluid stiffness depletion to be described by Eq.~(\ref{eqn:delta_ns_final}). All parameters entering Eq.~(\ref{eqn:delta_ns_final}) can be straightforwardly simulated either through direct Monte Carlo estimators or through numerical derivatives. For $U/t=4$ 
and filling factor $n\approx0.693$ we find compressibility $\varkappa=0.2974(5)t$ and sound velocity $c=2.087(3)ta$. Numerical derivatives yield the parameters $\nu=1.763(5)ta^2$ and $\gamma=-0.10(7)ta^4$. The most difficult parameter to estimate is $\sigma$, for which we find $\sigma=-0.15(10)t$. The fit-free theoretical prediction with $\sigma=-0.1t$ is compared to numerical results for $n_s$ in Fig.~\ref{fig:nsT3soft}. The 
fit-free theoretical prediction falls perfectly within the statistical 
error bars of computed $n_s(T)$ points.

\textit{Universal thermodynamics.} As mentioned above, the main advantage of the hydrodynamic Hamiltonian formalism is that it allows us to work in the canonical ensemble by systematically identifying and discarding terms originating from the temperature-induced change in total density. The problem of calculating depletion in the grand canonical ensemble is simpler, and arguably more illuminating. The canonical depletion $\Delta n_s(n, T)$ 
follows from the grand canonical depletion $\Delta n_s(\mu, T)$ 
by a simple relation
\begin{equation}\label{eqn:ns_n_relation}
    \Delta n_s(n,T) = \Delta n_s(\mu,T)-\nu \Delta n(\mu,T) \, ,
\end{equation}
where $\Delta n(\mu, T)$ is the equation of state.

In the absence of constraints on the total density, we can work directly with the phase-only version of Popov's hydrodynamic action, defined by the Lagrangian density
\begin{equation}\label{eqn:phase_lagrangian}
    \begin{split}
    \mathcal{L}(\Phi) = &-\frac{n_s^{(0)}}{2}(\nabla\Phi)^2-\frac{\varkappa}{2}\dot{\Phi}^2-\frac{i\varkappa\nu}{2}\dot{\Phi}(\nabla\Phi)^2 \\
    &+\frac{\varkappa^2\tilde{\gamma}}{2}\dot{\Phi}^2(\nabla\Phi)^2-\frac{\tilde{\sigma}}{4}(\nabla\Phi)^4\,,
    \end{split}
\end{equation}
where
\begin{equation}
    \tilde{\gamma}=\gamma-\frac{\varkappa\nu\lambda}{2},\qquad\tilde{\sigma}=\sigma-\frac{\varkappa\nu^2}{2}\,,
\end{equation}
and
\be\label{eqn:lambda_derivative}
\lambda \, =\, \frac{\partial^3 \mathcal{E}(n,k_0^2)}{\partial n^3}\bigg|_{k_0=0} \, .
\ee
Perfoming decomposition~(\ref{eqn:phi_decomp}) and keeping only bilinear terms in the Lagrangian~(\ref{eqn:phase_lagrangian}) allows one to calculate the partition function explicitly in Fourier space as a function of $k_0$. Thermodynamic quantities are generated by taking derivatives of the grand potential density, which takes the form
\begin{equation}\label{eqn:grand_potential}
\frac{1}{V}\Omega(\mu, T, k_0) = \frac{T}{2V}\sum_{\mathbf{k},\omega}\ln \epsilon(\mathbf{k},\omega) \, ,
\end{equation}
where
\begin{equation}\label{eqn:epsilon_def}
    \begin{split}
    \epsilon(\mathbf{k},\omega)=\pi^{-1}[&(n_s^{(0)}+\tilde{\sigma}k_0^2)k^2+(\varkappa-\varkappa^2\tilde{\gamma}k_0^2)\omega^2 \\
    & +2i\varkappa\nu k_0\omega k_x + 2\tilde{\sigma}k_0^2k_x^2]
    \end{split}
\end{equation}
is a bi-linear function of the Fourier wavevector $\mathbf{k}$ and bosonic Matsubara frequency $\omega$. Upon proper UV-regularization of Eq.~(\ref{eqn:grand_potential}), 
it generates $\Delta n_s(\mu,T)$  and $\Delta n(\mu, T)$ through derivatives over
$k_0$ and $\mu$. 
The bi-linear structure of $\epsilon(\mathbf{k}, \omega)$ implies
universal $T^{d+1}$ scaling (in an infinitely large system) of $\Omega$, and,
correspondingly, the thermodynamic functions generated by $\Omega$.

\textit{Helium in disordered media.} Multiple studies have been performed on superfluid density depletion in disordered media: \he4 in compressed lampblack and Vycor~\cite{reppy1972,reppy1975sound, reppy1986torsion} and \hel3 in aerogel~\cite{halperin2018aerogel}. Thus, it is natural to ask whether the experimental findings are consistent with (or relevant to) our asymptotic theory. The data for \hel3 in aerogel are not sufficient to clearly resolve the low-temperature regime and therefore cannot be directly compared with our results. 

The low-temperature data for \he4 do exist, but are quite controversial. 
In compressed lampblack \cite{reppy1972}, the depletion measured by two different techniques---via temperature dependence of the frequency of U-tube and fourth-sound oscillations---was found to be linear in $T$. In Vycor, it was found \cite{reppy1975sound} that the finite-temperature correction to the frequency of the fourth-sound mode scaled as $T^2$, in contrast to $T$ found in lampblack and $T^4$ predicted by our theory. 

\begin{figure}
    \centering
    \includegraphics[width=0.9\linewidth]{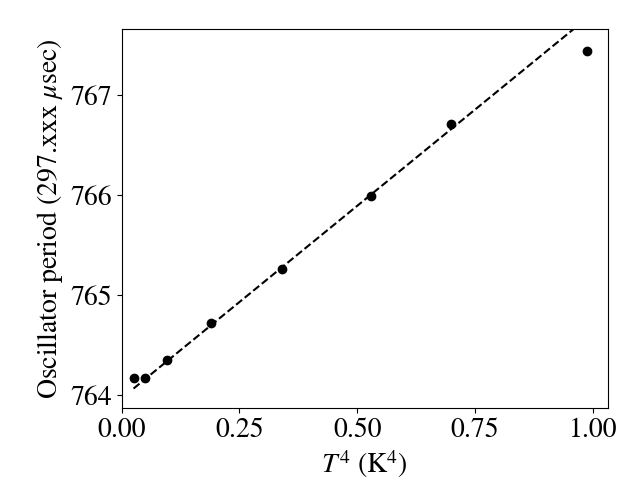}
    \caption{Period of torsional oscillator as a function of $T^4$ for liquid \he4 in Vycor. Data was digitized from the 1.08 bar curve in Fig.~11 in Ref.~\cite{reppy1986torsion}. Dashed line is a fit to a $T^4$ power law. The data was obtained in the experimental setup  consisting of a torsional oscillator containing a cylinder of Vycor glass with \he4 admitted to the Vycor through a narrow tube connected to a reservoir of bulk helium, kept at a constant pressure.}
    \label{fig:reppy_data}
\end{figure}

Finally, an almost perfect $T^4$ scaling was observed for \he4 in Vycor,
see Fig.~\ref{fig:reppy_data} showing low-temperature data from Ref.~\cite{reppy1986torsion} in the torsional oscillator setup at constant pressure. The change of the superfluid density in Vycor is due to two different effects, both, however, supposed to demonstrate the same $T^4$ low-temperature scaling: (i) the finite-temperature depletion at fixed density and (ii) the ground-state change of the superfluid density due to the change in the chemical potential (induced by the change of the temperature of the bulk helium at fixed pressure). Unfortunately, measured temperatures are not sufficiently low to claim the asymptotic regime. Nevertheless, the absence of visible $T$ and $T^2$ corrections within the given temperature interval suggests that these terms are negligible or absent altogether.

\textit{Concluding remarks.} In a Galilean system, Landau theory relates
the depletion of the superfluid stiffness to the wind of elementary exitations. 
However, a na\"ive  attempt to apply Landau theory to systems with broken Galilean symmetry immediately fails. The reason becomes clear by studying Eq.~(\ref{eqn:delta_ns_final}). Landau's theory of the phonon wind is fully captured by the parameter $\nu$, originating from the cubic term in the hydrodynamic Hamiltonian~(\ref{eqn:H_hydro_0}). To correctly describe the leading contribution to the depletion, one must also include the quartic terms in the second line of Eq.~(\ref{eqn:H_hydro_0}). In fact, there exist systems with certain symmetries, for instance particle-hole symmetry, where $\nu$ is identically equal to zero. In this case, the depletion is controlled entirely by the parameters $\gamma$ and $\sigma$ fundamentally absent in Landau theory. The particle-hole symmetry may be realized by taking $U\rightarrow\infty$ in the Hamiltonian~(\ref{eqn:bose_hubbard}) while keeping $\mu=0$ (the \textit{hard-core boson} model). Superfluid stiffness depletion in this model has been thoroughly explored in the companion paper \cite{companion}.

The other issue addressed in Ref.~\cite{companion} is interplay between finite-size and finite-temperature effects on the depletion of superfluid stiffness. At harmonic level, inverse temperature plays the role of the $(d+1)$-st spatial direction in the theory. This implies, in addition to the universal $T^{d+1}$ scaling, a universal $1/L^{d+1}$ scaling of $\Delta n_s$ in systems at $T=0$ with slab geometries. 

A large portion of existing experimental data on the low-temperature depletion of $\rho_s$ in \he4 in three-dimensional disordered media is profoundly inconsistent with both the universal $T^4$ scaling and with each other---suggesting either linear or quadratic laws. Our asymptotically exact theory calls for further experimental studies towards resolving the controversy. Among the reasons why the low-temperature experiment with \he4 in disordered media did not demonstrate the universal behavior found in our work we would like to emphasize (i) potential lack of adiabaticity of the response of the superfluid to torsional oscillations and (ii) possible effects of long-range-correlated disorder. Extensive numeric simulations of low-temperature disordered superfluids might shed a direct light on the relevance of the latter reason.

\begin{acknowledgments}
This work was stimulated by numerous conversations (though without reaching a consensus \cite{hegg2024universallowtemperaturefluctuationunconventional}) with Wei Ku about the scientific status of Landau theory. The discussion of experiments in helium was inspired by an exchange with Grigory Volovik.
This work was supported by the National Science Foundation under Grant DMR-2335904.
\end{acknowledgments}

\bibliography{refs.bib}

\end{document}